\documentclass[letterpaper, 10 pt, conference]{ieeeconf}%

\IEEEoverridecommandlockouts                              %

\usepackage{xr}

\newcommand{\omitted}[1]{}%

\title{%
Distributed Equilibrium-Seeking in Target Coverage Games via Self-Configurable Networks under Limited Communication}
\author{Jayanth Bhargav$^{1*}$, Zirui Xu$^{2*}$, Vasileios Tzoumas$^{2\dagger}$,  Mahsa Ghasemi$^{1\dagger}$, Shreyas Sundaram$^{1\dagger}$
\thanks{$^*$These authors contributed equally to this work.}
\thanks{$^\dagger$These authors contributed equally to this work.}
	\thanks{%
    $^1$Elmore Family School of Electrical and Computer Engineering, Purdue University, West Lafayette, IN 47907 USA;  {\tt\footnotesize \{jbhargav,mahsa, sundara2\}@purdue.edu} }
 \thanks{$^2$Department of Aerospace Engineering, University of Michigan, Ann Arbor, MI 48109 USA;  {\tt\footnotesize \{ziruixu,vtzoumas\}@umich.edu}} 
    \thanks{This work was supported in part by the National Science Foundation (NSF) CAREER Award No. 2337412, the Army Research Office (ARO) Early Career Program Award W911NF-25-1-0280, the Office of Naval Research (ONR) and Saab, Inc. under the Threat and Situational Understanding of Networked Online Machine Intelligence (TSUNOMI) program (grant no. N00014-23-C-1016) and the National Aeronautics and Space Administration (NASA) under grant 80NSSC24M0070. Any opinions, findings, and conclusions or recommendations expressed in this material are those of the author(s) and do not necessarily reflect the views of the NSF, ARO, ONR, Saab, Inc., and/or NASA.}}

\usepackage{cite}

\usepackage{comment}
\usepackage{siunitx}
\usepackage{relsize}
\usepackage{ifthen}
\usepackage[colorinlistoftodos]{todonotes}

\usepackage[vlined,ruled,linesnumbered]{algorithm2e}
\usepackage{graphics} %
\usepackage{rotating}
\usepackage{color}
\usepackage{enumerate}
\usepackage[T1]{fontenc}
\usepackage{psfrag}
\usepackage{epsfig} %
\usepackage{booktabs}
\usepackage{graphicx,url}
\usepackage{multirow}
\usepackage{array}
\usepackage{latexsym}
\usepackage{amsfonts}
\usepackage{amsmath}
\usepackage{amssymb}
\usepackage{xstring}
\usepackage{algorithmic}
\usepackage{multirow}
\usepackage{xcolor}
\usepackage{prettyref}
\usepackage{flexisym}
\usepackage{bigdelim}
\usepackage{breqn} %
\usepackage{listings}
\let\labelindent\relax
\usepackage{xspace}
\usepackage{bm}
\graphicspath{{./figures/}}
\usepackage{tikz}
\usetikzlibrary{matrix,calc}
\usepackage{lipsum}
\usepackage{mdwlist}

\let\itemize\undefined
\makecompactlist{itemize}{stditemize}
\usepackage{enumitem}
\usepackage{caption}

\usepackage{amsthm}
\usepackage{mathtools}

\makeatletter
\let\NAT@parse\undefined
\makeatother
\usepackage{hyperref}
\usepackage{cleveref}

\hypersetup{%
  colorlinks=true,
  linkcolor=blue,
  filecolor=magenta,      
  urlcolor=black,
  citecolor=red,
  linkbordercolor={0 0 1}
}

\newtheorem{theorem}{Theorem}
\newtheorem{problem}{Problem}

\newtheorem{lemma}{Lemma}

\newtheorem{definition}{Definition}

\newcommand{\bdmath}{\begin{dmath}}
\newcommand{\edmath}{\end{dmath}}
\newcommand{\beq}{\begin{equation}}
\newcommand{\eeq}{\end{equation}}
\newcommand{\bdm}{\begin{displaymath}}
\newcommand{\edm}{\end{displaymath}}
\newcommand{\bea}{\begin{eqnarray}}
\newcommand{\eea}{\end{eqnarray}}
\newcommand{\beal}{\beq \begin{array}{lll}}
\newcommand{\eeal}{\end{array} \eeq}
\newcommand{\beas}{\begin{eqnarray*}}
\newcommand{\eeas}{\end{eqnarray*}}
\newcommand{\ba}{\begin{array}}
\newcommand{\ea}{\end{array}}
\newcommand{\bit}{\begin{itemize}}
\newcommand{\eit}{\end{itemize}}
\newcommand{\ben}{\begin{enumerate}}
\newcommand{\een}{\end{enumerate}}

\newcommand{\calA}{{\cal A}}
\newcommand{\calB}{{\cal B}}
\newcommand{\calC}{{\cal C}}

\newcommand{\calE}{{\cal E}}

\newcommand{\calG}{{\cal G}}

\newcommand{\calM}{{\cal M}}
\newcommand{\calN}{{\cal N}}

\newcommand{\calR}{{\cal R}}

\newcommand{\calV}{{\cal V}}

\definecolor{myblue}{RGB}{65 105 225}

\newcommand{\hide}[1]{}

\newcommand{\hiddenText}{{\color{gray} hidden text.}}
\newcommand{\hideWithText}[1]{\hiddenText}

\DeclareMathOperator*{\argmax}{arg\,max}

\newcommand{\opt}{^{\star}}

\DeclareRobustCommand{\scenario}[1]{%
  \ifmmode
    \mathsf{#1}%
  \else
    \texorpdfstring{{\fontsize{8.9}{9}\selectfont\sffamily #1}\xspace}{#1}%
  \fi
}

\newcommand{\ie}{\emph{i.e.},\xspace}
\newcommand{\eg}{\emph{e.g.},\xspace}
\newcommand{\myin}{\, \in \,}

\newcommand{\myParagraph}[1]{{\bf #1.}\xspace}

\renewcommand{\opt}{\scenario{OPT}}

\newcommand{\curv}{\kappa}

\newcommand{\actionsel}{\scenario{ActSel}}
\newcommand{\neighborsel}{\scenario{NeiSel}}

\newcommand{\elem}{v}

\newcommand{\distfsf}{p}

\newcommand{\solopt}{\calA^{\opt}}

\newcommand{\expthree}{\scenario{EXP3}}

\newcommand{\smi}[2]{\text{{\fontsize{9}{9}\selectfont\sf VoC}}_{f_t,t}({#1};\,{#2})}
\newcommand{\voc}{\scenario{VoC}}

\PassOptionsToPackage{end}{algorithmic}

\begin{document}

\maketitle
\thispagestyle{empty}
\pagestyle{empty}

\begin{abstract}
We study a target coverage problem in which a team of sensing agents, operating under limited communication, must collaboratively monitor targets that may be adaptively repositioned by an attacker. 
We model this interaction as a zero-sum game between the sensing team (known as the defender) and the attacker. However, computing an exact Nash equilibrium (NE) for this game is computationally prohibitive as the action space of the defender grows exponentially with the number of sensors and their possible orientations. Exploiting the submodularity property of the game's utility function, we 
propose a distributed framework that enables agents to self-configure their communication neighborhoods under bandwidth constraints and collaboratively maximize the target coverage. We establish theoretical guarantees showing that the resulting sensing strategies converge to an approximate NE of the game. To our knowledge, this is the first distributed, communication-aware approach that scales effectively for games with combinatorial action spaces while explicitly incorporating communication constraints. To this end, we leverage the distributed bandit-submodular optimization framework and the notion of \textit{Value of Coordination} that were introduced in~\cite{xu2026self}. Through simulations, we show that our approach attains near-optimal game value and higher target coverage compared to baselines.
\end{abstract}


\section{Introduction}\label{sec:Intro}

In today’s evolving landscape of security and surveillance, effective deployment of sensors is critical to ensure comprehensive monitoring and target coverage~\cite{guvensan2011coverage}. Game-theoretic frameworks are widely used to model adversarial sensing and coverage problems. Many existing formulations model the coordination among sensing agents through a centralized decision-making process~\cite{bhargav2025sensor}. However, in practice, large surveillance areas and limited communication make it infeasible to have a central coordinator~\cite{xu2023bandit, atanasov2015decentralized, corah2018distributed}. These challenges instead necessitate distributed coordination, in which agents selectively communicate with local neighbors and adapt their local actions to maximize overall team performance.

In this work, we consider a target coverage task where a team of sensing agents (defender) must coordinate to cover targets deployed in an environment. The utility function of the defender, for a given target deployment by the attacker, is the fraction of targets the agents collectively cover. This utility function is submodular---a property that captures diminishing returns due to overlapping coverage among agents~\cite{fisher1978analysis}. This structure enables the use of efficient approximation algorithms for utility maximization~\cite{bhargav2024submodular}. However, leveraging submodularity becomes more challenging when the utility value changes over time due to the attacker adaptively repositioning the targets.
\begin{figure}[!t]
    \centering
    \captionsetup{font=footnotesize}
    \includegraphics[width=\linewidth]{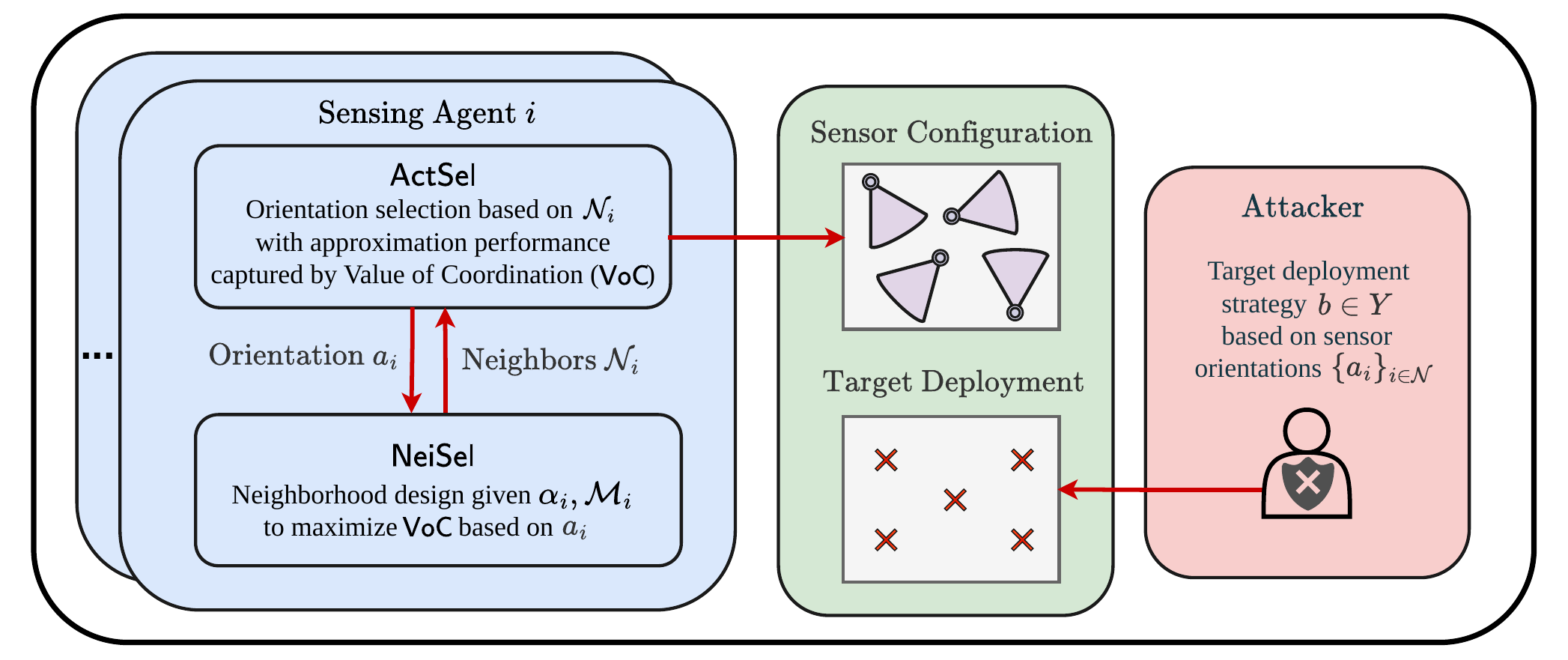}
    \caption{\textbf{Distributed Target Coverage Game.} Team of sensing agents choose communication neighbors and orientations to maximize target coverage. Simultaneously, the attacker selects a target deployment strategy to achieve the least coverage for the sensing agents. The algorithms {\fontsize{7}{7}\selectfont\sffamily ActSel}, {\fontsize{7}{7}\selectfont\sffamily NeiSel}, and their interaction framework via the \textit{Value of Coordination}, follows~\cite{xu2026self}.}
    \label{fig:game}
    \vspace{-5mm}
\end{figure}
 We model this interaction as a zero-sum game between the sensing team and an adaptive attacker. Then, at Nash equilibrium (NE), neither the sensing team nor the attacker can improve their objective through unilateral deviation, yielding sensing strategies that are robust to such strategic adaptations. Solving this game for the exact NE strategies, however, can become computationally intractable due to the combinatorial growth of the defender’s action space with the number of sensors and their orientations, as well as the need to propagate global information without a central coordinator. To address this, we propose a distributed, scalable and communication-efficient framework for approximately solving this game. Our approach draws on and unifies ideas from online learning, game theory, and distributed bandit-submodular optimization. 

\textbf{Related Work.} Several recent works study large zero-sum games, including analyses of equilibrium dynamics~\cite{konda2025best} and algorithms for estimating equilibria~\cite{huang2024no,liao2025distributed, wilder2018equilibrium,li2021cfr}. In~\cite{huang2024no,liao2025distributed}, the authors study subnetwork games with Lipschitz continuous utilities and require a strongly connected network to ensure convergence to an NE. In~\cite{wilder2018equilibrium}, the authors present a multilinear extension--based technique with an approximation guarantee in a centralized setting. On the other hand, the regret minimization approach of~\cite{li2021cfr} does not provide guarantees on the quality of the resulting strategies. 
In contrast to these works, we establish near-optimality guarantees in a distributed setting without requiring connectivity assumptions on the communication network.

Related work also focuses on solving distributed submodular optimization problems in real-time, which are challenging since such problems are known to be NP-hard~\cite{Feige:1998:TLN:285055.285059}. Although existing algorithms achieve near-optimal solutions in polynomial time, they often need significant runtime to terminate in distributed settings due to limited communication speeds and bandwidths~\cite{atanasov2015decentralized,robey2021optimal,konda2022execution}. In distributed settings, multiple works have focused on online submodular maximization in unknown (dynamic) environments~\cite{streeter2008online,xu2023bandit}. They study these in a bandit feedback setting, and provide bounded suboptimality guarantees with respect to optimal time-(in)variant actions through regret minimization~\cite{auer2002nonstochastic}. However, these works (i) can require extensive communication rounds as in~\cite{atanasov2015decentralized,konda2022execution,robey2021optimal}, and (ii) require a connected network for information propagation, which is not always feasible~\cite{wu2021comprehensive}. 

\textbf{Contributions.} We propose a distributed framework for a submodular target coverage game where a set of sensing agents coordinate their orientations to maximize the target coverage, while the attacker adaptively relocates the deployed targets to minimize the coverage (\Cref{alg:main}). Our framework enables a near-optimal coordination strategy for sensing agents that simultaneously configure communication neighborhoods under limited bandwidth budgets and coordinate orientations, assuming a no-regret attacker that can learn the best response. To this end, we leverage the distributed bandit-submodular optimization framework in~\cite{xu2026self} and extend it to the game-theoretic setting. The framework can asymptotically achieve an approximate NE of the game. During the repeated play of the game, sensing agents alternate between selecting their orientations and neighbors to adapt to the time-varying target deployments by the attacker. 
In particular, our framework has the following key properties.
\paragraph{Performance Guarantees}  
It enables the distributed team of agents to asymptotically achieve an approximate NE. %
Despite the communication bandwidth constraints, the agents achieve near-optimal performance by each actively adapting the neighborhood over time, enabled by optimizing its Value of Coordination (\voc), a quantity introduced in~\cite{xu2026self}. 
\paragraph{Anytime Self-Configuration}  
Our framework enables each agent to choose its actions and neighbors individually using local information, enabling seamless adaptation to near-optimal orientations and neighborhoods as agents join or leave the network (\eg sensor failures or additions).

Finally, through simulations, we demonstrate that the sensing agents achieve game utility arbitrarily close to the NE value and higher target coverage compared to baselines.

\section{Problem Formulation}\label{sec:problem}

In this section, we formalize the \textit{Distributed Adversarial Target Coverage Game} and lay down the framework about  the defender,  attacker and game utility. To this end, we use the notation: $\calV_\calN \triangleq \prod_{i\myin \calN} \,\calV_i$ is the cross product of sets $\{\calV_i\}_{i\myin \calN}$; %
$[T]\triangleq\{1,\dots,T\}$ for any positive integer $T$; $f(a\,|\,\calA)\triangleq f(\calA \,\cup\, \{a\})-f(\calA)$ is the marginal gain of set function $f:2^\calV\mapsto \mathbb{R}$ for adding $a \in \calV$ to $\calA \subseteq\calV$; and $|\calA|$ is the cardinality of a discrete set $\calA$. 

\noindent \myParagraph{Communication network} The communication network is denoted by $\calG=(\calN, \calE)$, where $\calN$ is the set of nodes/agents and $\calE$ refers to edges between agents. $\calE$ is \textit{unspecified} a priori; instead, the agents must form and optimize this communication network internally to coordinate their orientation actions and maximize their joint utility.

\noindent \myParagraph{Communication neighborhood}  
When a communication channel exists from agent $j$ to $i$, \ie $(j\rightarrow i) \in \calE$, then $i$ can receive, store, and process information from $j$.  The set of all agents that $i$ receives information from is denoted by $\calN_i$---agent $i$'s \textit{neighborhood}. When we say that an agent exchanges \textit{information}, we mean that it communicates the set of targets it currently covers under its chosen orientation 
(\eg by encoding and sharing unique identifiers for targets).

\noindent \myParagraph{Communication constraints} Each agent $i$ can receive information from up to $\alpha_i$ other agents due to onboard bandwidth constraints. Thus, it must be $|\calN_i|\leq \alpha_i$. Also, we denote by $\calM_i$ the set of agents that have agent $i$ within reach -- not all agents may have agent $i$ within reach because of distance or obstacles. Therefore, agent $i$ can pick its neighbors by choosing at most $\alpha_i$ agents from $\calM_i$. Evidently, $\calN_i\subseteq\calM_i$.

\noindent \myParagraph{Defender's action space}
For each sensing agent $i \in \mathcal{N}$, we denote the agent's sensing action space by the set of feasible orientations $\mathcal{V}_i$. An action $a_i \in \mathcal{V}_i$ corresponds to a specific orientation of sensor $i$. The joint action of the sensing team is given by $ \calA = (a_1, \ldots, a_{|\mathcal{N}|}) \in \mathcal{V}_{\mathcal{N}}$. We denote a mixed strategy of the defender by $x \in \Delta_{|X|}$, where $X = \mathcal{V}_{\mathcal{N}}$ is the set of joint actions, and $\Delta$ denotes a probability simplex.

\noindent \myParagraph{Attacker's action space} The attacker places a fixed number of targets according to one of finitely many deployment configurations \( Y = \{b_1, \ldots, b_m\} \), where each \( b_i \in Y \) is a distinct spatial arrangement/deployment of these targets in the environment. We denote the mixed-strategy of the attacker by $y \in \Delta_{|Y|}$, where $\Delta$ denotes a probability simplex.

\noindent \myParagraph{Game Utility} The game utility $f(\calA, b)$ for a pure-strategy pair $(\calA, b)$, is the fraction of targets covered by the sensing team with the joint orientation $\calA$ under the target deployment configuration $b$. The team of sensing agents aim to maximize~$f$, while the attacker aims to minimize $f$, which results in a zero-sum game. We denote the game matrix by $G \in \mathbb{R}^{|X| \times |Y|}$, where $G_{ij} = f(\calA^i, b_j), \calA^i \in X, b_j \in Y$.

With the notation introduced above, we now formalize the \textit{Distributed Target Coverage Game} below.

\begin{problem}[Distributed Target Coverage Game]
\label{pr:game}
The zero-sum game between the team of sensing agents and the attacker is given by the following optimization problem.
\begin{equation}
    \label{eq:game}
   \max_{x \in \Delta_{|X|}}
    \min_{y \in \Delta_{|Y|}} x^{\top}Gy =  \max_{x \in \Delta_{|X|}}
    \min_{y \in \Delta_{|Y|}}
    \mathbb{E}_{\calA \sim x, b \sim y}\, f(\calA, b).
\end{equation} 
\end{problem}

The solution to the zero-sum game in  Problem \ref{pr:game}, i.e., the NE strategies, can be obtained by solving~\cref{eq:game}. As stated earlier, exact NE computation is impractical due to the combinatorial action space of the defender (\eg $10$ sensors with $4$ orientations yield $4^{10}$ joint actions) and the absence of a central coordinator. This motivates the design of an efficient, distributed  algorithm that relies only on local coordination between the agents and achieves near-optimal performance relative to the centralized NE. In zero-sum games, near-optimality is formalized through the notion of an $\epsilon$-Nash equilibrium.

\begin{definition}[$\epsilon$-Nash Equilibrium ($\epsilon$-NE)~\cite{bhargav2025sensor}]
A pair of mixed strategies $(x^\star, y^\star)$ is an \textit{$\epsilon$-Nash equilibrium} of the zero-sum game with payoff matrix $G$ if $x^{\star \top} G y^\star \ge x^\top G y^\star - \epsilon,
    \; \forall x \in \Delta_{|X|}$
and $x^{\star \top} G y^\star \le x^{\star \top} G y + \epsilon,
    \; \forall y \in \Delta_{|Y|}.$
\end{definition}

In other words, an $\epsilon$-NE ensures that neither player can improve their payoff by more than $\epsilon$ relative to the game value (NE value) through unilateral deviation.

\vspace{-1mm}\section{Scalable Equilibrium Approximation}\label{sec:algorithm}

In this section, we introduce our main algorithm which estimates an $\epsilon$-NE of the Distributed Target Coverage Game. We begin by characterizing key structural properties of the game’s utility function.

 \subsection{Preliminaries}
\begin{definition}[Normalized and Non-Decreasing Submodular Set Function{~\cite{fisher1978analysis}}]\label{def:submodular}
A set function $g:2^\calV\mapsto \mathbb{R}$ is \emph{normalized and non-decreasing submodular} if and only if 
\begin{itemize}
\item (Normalization) $g(\,\emptyset\,)=0$;
\item (Monotonicity) $g(\,\calA\,)\leq g(\,\calB\,)$, $\forall\,\calA\subseteq \calB\subseteq \calV$;
\item (Submodularity) $g(\,s\,|\,\calA\,)\geq g(\,s\,|\,{\mathcal{B}}\,)$, $\forall\,\calA\subseteq {\mathcal{B}}\subseteq\calV$, $s\in \calV$.
\end{itemize}
\end{definition}

\begin{definition}[2nd-order Submodularity{~\cite{foldes2005submodularity}}]\label{def:conditioning}
A set function $g:2^\calV\mapsto \mathbb{R}$ is \emph{2nd-order submodular} if and only if 
\begin{equation}\label{eq:conditioning}
    g(s\,|\,\calC) - g(s\,|\,\calA\cup\calC) \geq g(s\,|\,\calB\cup\calC) - g(s\,|\,\calA\cup\calB\cup\calC),
\end{equation}
for any \emph{disjoint} $\calA, \calB, \calC\subseteq \calV$ ($\calA \cap \calB \cap \calC =\emptyset$) and  $s\in\calV$.
\end{definition}

For a target deployment configuration $b \in Y$, we define $f(a\,|\,\calA, b)\triangleq f(\calA \cup \{a\}, b)-f(\calA, b)$ to be the marginal gain of the game utility function for adding $a \in \calV$ to $\calA \subseteq\calV$, given $b \in Y$. Then, for a given $b \in Y$, the game's utility function $f(\cdot, b)$ is normalized, monotone, submodular, and 2nd-order submodular over the set of sensing actions (sensor orientations), with submodularity arising from overlapping coverage areas among the sensing agents~\cite{corah2018distributed}. We leverage these properties to guide our algorithm design.

\setlength{\textfloatsep}{3mm}
\begin{algorithm}[!b]
	\caption{Repeated Play Dynamics of the Distributed Target Coverage Game 
	}
	\begin{algorithmic}[1]
		\REQUIRE \!Number of time steps $T$; neighbor candidate sets $\calM_i$, neighborhood sizes $\alpha_i$, and action sets $\calV_i$ for all sensing agent $i \in \calN$; attacker's action set $Y$; game's utility function $f$. %
		\ENSURE A pair of $\epsilon$-NE strategies $(\bar{x}_T, \bar{y}_T)$.
		\medskip
        \STATE $\calN_{i,0}\gets\emptyset, \forall i \in \calN$; 
        \STATE $\eta^b\gets\sqrt{2\log{|Y|}\,/\,(|Y|T)}$;
            \STATE $w_{1} \gets\left[w_{1,1}, \dots, w_{|Y|,1}\right]^\top$ with $w_{b,1}=1, \forall b \in Y$;
            \STATE $q_{1}\gets{w_{1}}\,/\,{\|w_{1}\|_1}$;
		\FOR{each time step $t\in [T]$}
        \STATE \textbf{sample} target deployment config. $b_{t}\in Y$ \textbf{from} $q_{t}$;
        \STATE $p_{i,t}, a_{i,t}\gets\text{\actionsel}(b_{1:t-1}, T, \calV_i, f) , \forall i \in \calN$;
        \STATE $\calN_{i,t}\gets\text{\neighborsel}(a_{i,t}, b_{1:t-1}, T, \calM_i, \alpha_i, f), \forall i \in \calN$;
        \STATE \textbf{receive} neighbors' actions $\{a_{j, t}\}_{j\myin\calN_{i,t}}, \forall i \in \calN$;
        \STATE \textbf{update} \actionsel (per lines 6--8 of~Algorithm \ref{alg:action}) \& \neighborsel (per lines 6--11 of~Algorithm \ref{alg:neighbor});
        \STATE $r_{b, t}\gets f(\{a_{i, t}\}_{i\in\calN},b_t)$ 
        
            \STATE $\hat{r}_{b,\,t} \gets 1 - \frac{{\bf 1}(b_{t}\,=\,b)}{p_{b,t}}\left(1\,-\,r_{b, t}\right), \forall b \in Y$;
            \STATE $w_{b,t+1}\gets w_{b,t}\exp{(-\eta^b \,\hat{r}_{b,t})}$;	
            \STATE \textbf{get} distribution $q_{t+1}\gets{w_{t+1}}\,/\,{\|w_{t+1}\|_1}$;
		\ENDFOR
        \STATE \textbf{estimate} joint-strategy $\bar{p}_{t} \triangleq \Pi_{i\in\calN}\, p_{i,t}, \forall t \in [T]$.
        \STATE $\bar{x}_T \gets \frac{1}{T} \sum_{t \in [T]} \bar{p}_{t},\; \bar{y}_T \gets \frac{1}{T} \sum_{t \in [T]} q_{t}$.
	\end{algorithmic}\label{alg:main}
\end{algorithm}

\subsection{Main Algorithm}
Algorithm~\ref{alg:main} outlines the iterative interaction between the attacker and the defender. Following standard game-theoretic modeling, we assume that the attacker is a rational agent that follows a no-regret learning strategy~\cite{farina2017regret}. Such learning dynamics are known to converge to NE in zero-sum games~\cite{daskalakis2019last}. Under this modeling, the interaction between the sensing agents and the attacker induces an iterative game-play dynamic. We show that, within this dynamic, the decision-making problems faced by the attacker and each sensing agent can be naturally formulated as adversarial multi-armed bandits (A-MAB)~\cite{auer2002nonstochastic}, because the agents and the attacker are unaware of each other's actions a priori. 

From the defender’s perspective, at each round the attacker selects a target deployment configuration, and the sensing team must choose an orientation strategy to maximize target coverage. When the attacker adaptively changes the deployment over time, the sensing team must learn an orientation-selection policy that maximizes the expected coverage utility. This setting corresponds to an A-MAB problem, where each arm represents an orientation strategy and the reward at each round is a function of the game’s utility based on the target deployment strategy chosen by the attacker in that round.

Similarly, from the attacker’s perspective, the sensing team’s adaptive orientation choices induce an adversarially evolving environment. The attacker then seeks a target deployment policy that minimizes the expected coverage achieved by the sensing team. This also constitutes an A-MAB problem, where the arms correspond to target deployment configurations and the reward is defined by the negative of game’s utility function, which the attacker is trying to maximize. The attacker follows a classical no-regret learning strategy \expthree~\cite{auer2002nonstochastic} by updating a distribution over target deployment strategies against a sequence of defender actions.

The defending team operates in a fully distributed manner, with each sensing agent maintaining two A-MAB instances: (i) \actionsel (Algorithm~\ref{alg:action}) - for selecting sensing actions; and (ii) \neighborsel (Algorithm~\ref{alg:neighbor})  - for selecting communication neighbors (see Appendix for details).
We show that this decomposition is highly beneficial, since under limited information exchange each agent can only evaluate the local utility, based on the subset of agents with which it communicates. Consequently, an agent’s utility is inherently determined by its communication neighborhood. As a result, choosing communication neighbors $\calN_{i,t}\subseteq\calM_i$ has a direct and non-trivial impact on the optimality of the sensing actions selected by each agent. This coupling makes neighbor selection an integral component of the overall decision process rather than a secondary design choice. To capture this formally, we leverage the notion of Value of Coordination. %

\begin{definition}[Value of Coordination (\voc)~\cite{xu2026self}]\label{def:MI}
Denote $f_t(\cdot)\triangleq f(\cdot, b_t)$. At any $t\in [T]$ with a target deployment $b_t \in Y$, for an agent $i\in\calN$ with an action $a_{i,t}$ and neighbors $\calN_{i,t}$, its Value of Coordination is defined as
\begin{equation}\label{eq:SubMI}
    \smi{a_{i,t}}{\calN_{i,t}} \triangleq f_t(a_{i,t}) - f_t(a_{i,t}\,|\,\{a_{j,t}\}_{j\in\calN_{i,t}}).
\end{equation}
\end{definition}

\voc~measures the overlap of utility between its action and its neighbors' actions, and per \cref{app-2}, the equilibrium approximation performance ($\calR_T^d$ as in \cref{eq:defender_regret}) will be tighter when \voc~is improved. Although \voc~is not computable a priori, according to~\cite[Lemma 1]{xu2026self}, it is normalized, non-decreasing, and submodular in $\calN_{i,t}$ given that $f$ is 2nd-order submodular (\Cref{def:conditioning}). Therefore, maximizing \voc~takes the form of bandit submodular maximization that can be solved by \neighborsel with bounded suboptimality (\Cref{alg:neighbor}).

\section{Performance Guarantees}\label{sec:tracking-regret}

We now establish performance guarantees for the proposed algorithm by characterizing bounds on the near-optimality of the resulting strategies relative to the NE of the game.

We first present our main theoretical result that extends~\cite[Theorem~1]{xu2026self}, which is for a fixed utility function $f$, by generalizing it to a time-varying setting in which the utility function value $f(\cdot, b_t)$ changes over time owing to attacker's actions.
Over $t\in [T]$, each agent $i\in\calN$ selects  actions $\{a_{i,t}\}_{t\in[T]}$ and neighborhoods $\{\calN_{i,t}\}_{t\in[T]}$ against a sequence of attacker's target deployments $\{b_t\}_{t\in [T]}$ (\Cref{alg:main}). Let $\solopt = \{a^{\opt}_1, \hdots, a^{\opt}_{|\calN|}\}$ denote the best joint action of the sensing team in hindsight. The cumulative regret of the defender against $\solopt$ is:
\begin{equation}
    \label{eq:defender_regret}
    \calR^d_T = \sum_{t=1}^T f(\solopt, b_t) - \sum_{t=1}^T f(\calA_t, b_t).
\end{equation}

\begin{definition}[Curvature~\cite{conforti1984submodular}]\label{def:curvature}
        The curvature of a normalized monotone submodular function $g\colon 2^{\calV}\mapsto \mathbb{R}$ is defined as
        \begin{equation}
            \kappa_g\triangleq 1-\min_{\elem\in\calV}{(g(\calV)-g(\calV\setminus\{\elem\}))}/{g(\elem)}.
        \end{equation}

\end{definition}$\kappa_g$ measures how far~$g$ is from modularity: if $\kappa_g=0$, then  $g(\calV)-g(\calV\setminus\{v\})=g(v)$, $\forall v\in\calV$, \ie $g$ is modular. Using this definition, we present our theoretical result.

\begin{theorem}[Defender's Average Regret Bound]\label{th:defender_regret} 
  After running \Cref{alg:main} for $T$ rounds, the defender achieves:
  \begin{align}
      \frac{\calR^d_T}{T} &\leq \kappa_f - \beta \, \rho(\kappa_{I}, \bar{\alpha}) \sum_{i\in\calN}\mathbb{E}_{t\sim T}\left[\scenario{VoC}\left(a_{i,t}; \calN_{i}^{\star}\right)\right] \nonumber\\
      & \quad+ \Tilde{O}\left(|\calN|\sqrt{\left(\bar{\alpha}^2\,|\bar{\calM}|\,+\,|\bar{\calV}|\right)/T}\right),
  \end{align}where $\kappa_{I}\triangleq \displaystyle \max_{i\in\calN}\kappa_{I,i}$, $\bar{\alpha}\triangleq \displaystyle  \max_{i\in\calN}\alpha_i$, $|\bar{\calV}|\triangleq\max_{i\in\calN}|\calV_i|$, $|\bar{\calM}|\triangleq \displaystyle \max_{i\in\calN}|\calM_i|$, $\kappa_f, \kappa_I \in[0,1]$, $\beta \triangleq \kappa_f\, (1-\kappa_f)$ and $\rho(\kappa_{I}, \bar{\alpha})\in[1-1/e, 1]$. Here $\kappa_f$ is the curvature of $f$, $\kappa_{I,i}$ is the curvature of \voc~of agent $i \in \calN$ and $\calN^{\star}_i \triangleq \argmax_{\calN_{i,t} \subseteq \calM_i; |\calN_{i,t}| \le \alpha_i} \sum_{t \in [T]} \smi{a_{i,t}}{\calN_{i,t}}$.
  
  \end{theorem}

Similarly, for a sequence of actions of the sensing agents $\{\calA_t=\{a_{1,t},\ldots,a_{|\calN|,t}\}\}_{t\in[T]}$, let $b^{\opt}$ denote the attacker strategy that is best in hindsight. The cumulative regret of the attacker with respect to this strategy is
\begin{equation}
\label{eq:attacker_regret}
\mathcal{R}^a_T
=
\sum_{t=1}^T f(\calA_t, b_t)
-
\sum_{t=1}^T f(\calA_t, b^{\opt}).
\end{equation}

Since the attacker follows the standard \expthree regime (lines 12--13 of \Cref{alg:main}), we have the following~\cite{lattimore2020bandit}.

\begin{lemma}[Attacker's Average Regret Bound]
  \label{lem:attacker_regret_bound}
  After running \Cref{alg:main} for $T$ rounds, the attacker achieves: 
  \begin{equation}
  \label{eq:attacker_regret_bound}
  \frac{\calR^a_T}{T} \leq \Tilde{O}\!\left(\sqrt{|Y|/{T}} \right).%
  \end{equation}
  \end{lemma}%

Combining the bounds from Lemma~\ref{lem:attacker_regret_bound} and Theorem~\ref{th:defender_regret} with \cite[Theorem 2]{farina2017regret} yields the following near-optimality guarantee for the strategies produced by Algorithm~\ref{alg:main}.

\begin{theorem}[Duality Gap]
  \label{thm:epsilon_ne}
  Let $(\bar{x}_T, \bar{y}_T)$ denote the strategies obtained after Algorithm \ref{alg:main} runs for $T$ rounds. Then, the strategies  $(\bar{x}_T, \bar{y}_T)$ are  $\epsilon_T$-NE of the Distributed Target Coverage Game, where
  \begin{align}
  \label{eq:epsilon_T}
      \epsilon_T &= \kappa_f - \beta \, \rho(\kappa_{I}, \bar{\alpha}) \sum_{i\in\calN} \displaystyle \mathbb{E}_{t \sim T}\left[\voc\left({{a_{i,t}; \calN^{\star}_{i,t}\}}}\right) \right] \nonumber\\
      &\quad+ \Tilde{O}\left(\sqrt{\left[|\calN|^2\left(\bar{\alpha}^2\,|\bar{\calM}|\,+\,|\bar{\calV}|\right) + |Y|\right]/T}\right).
  \end{align}
\end{theorem}
As $T \to \infty$, the limiting near-optimality parameter $\epsilon_\infty$ depends on $\kappa_f$ and the agents' \voc. By selectively coordinating with the most informative neighbors, we maximize \voc, thereby tightening the bound on $\epsilon_\infty$ by compensating for the suboptimality term $\kappa_f$.

\section{Numerical Evaluation}\label{sec:experiments}

\begin{figure}[!t]
    \captionsetup{font=footnotesize}
    \centering
    \vspace{1mm}
    \includegraphics[width=0.9\linewidth]{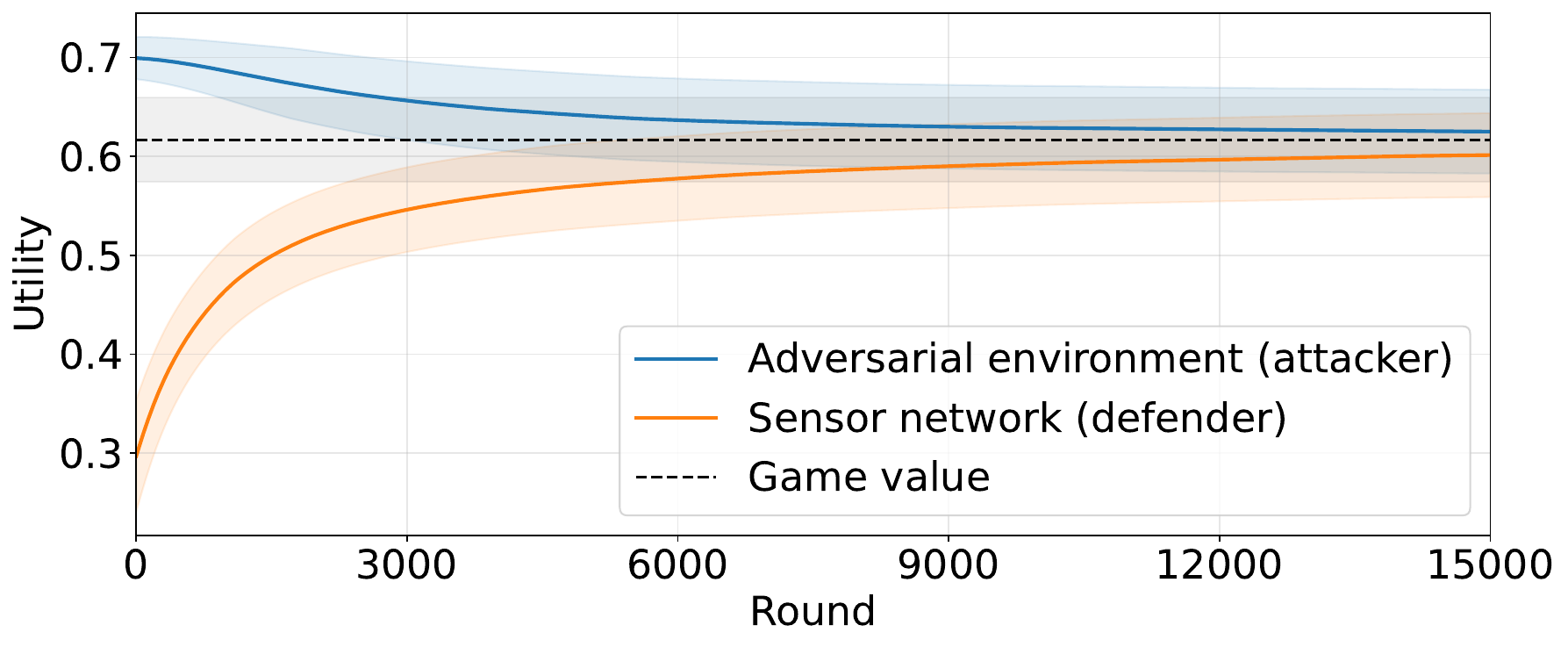}
    \caption{\textbf{Payoffs and duality gap.} Consider the repeated  game dynamics per Algorithm~\ref{alg:main} between three sensors each with bandwidth constraint $1$ (defender) and an $|Y| = 20$ possible target deployments (attacker). %
    As $T\rightarrow\infty$, the players asymptotically achieve the approximate NE, with the duality gap, \ie the difference between the players' payoffs, decreasing to its minimum, as shown in~\Cref{thm:epsilon_ne}. Results are averaged over 20 Monte Carlo trials, each with 15000 rounds.}
    \label{fig:duality}
\end{figure}

\begin{figure}[!b]
    \captionsetup{font=footnotesize}
    \centering
    \includegraphics[width=0.9\linewidth]{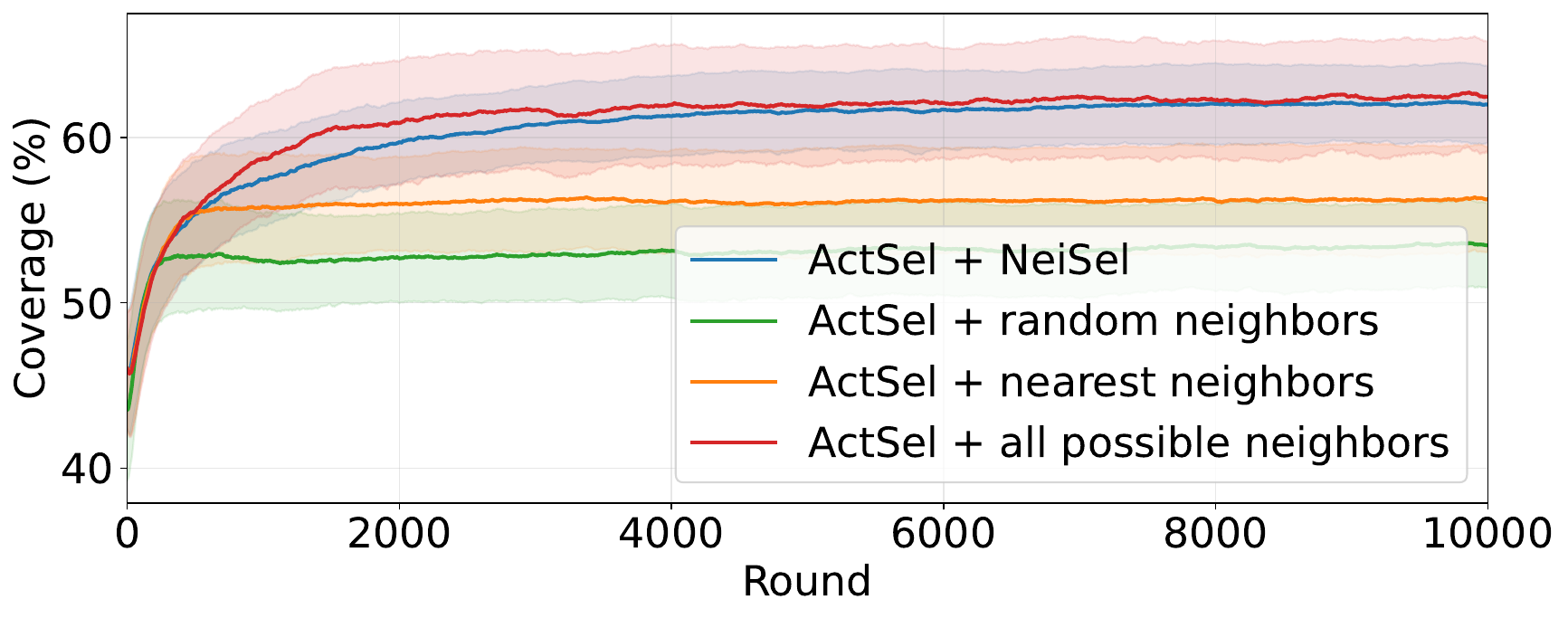}
    \caption{\textbf{Comparison of neighbor selection strategies.} Four algorithms are compared under the same action selection strategy ({\fontsize{7}{7}\selectfont\sffamily ActSel}) and adversarial environment ({\fontsize{7}{7}\selectfont\sffamily EXP3}), differing only in their neighbor selection strategies. 
    Results are averaged over 20 Monte Carlo trials, each with 10000 rounds.}
    \label{fig:coverage}
\end{figure}

Through two simulated scenarios of the target coverage game defined in~\Cref{pr:game}, we demonstrate that (i) Algorithm~\ref{alg:main} converges to  an approximate NE per \Cref{thm:epsilon_ne} (Figure~\ref{fig:duality}), 
and (ii) optimizing the network topology per Algorithm~\ref{alg:neighbor} leads to coverage performance that outperforms several baselines for neighborhood design, such as the nearest and random selection that are common in controls, %
and is even comparable to the best possible performance with unlimited bandwidth budgets (Figure~\ref{fig:coverage}). 

The experimental setup for both scenarios is as follows.
The sensors $\calN$ have fixed locations and need to each choose its orientation $a_{i,t}$ from the 16 cardinal directions $\calV_i$, $\forall t$. 
Each sensor $i$ is unaware of $a_{j,t}, j\in\calN\setminus\{i\}$; thus, they have to communicate to know about peers' actions. The attacker has a time-varying target distribution $b_t$ updated via \expthree (per Algorithm~\ref{alg:main}). 
 $f(\{a_{i,t}\}_{i\in \calN}, b_t)$ is the percentage of covered target area by the sensors when they select orientations $\{a_{i,t}\}_{i\in \calN}$ under the target distribution $b_t$. %
\paragraph{Convergence to Approximate NE of the Game}
We run Algorithm~\ref{alg:main} which simulates an iterative game play between the sensing agents (which alternate between \actionsel and \neighborsel) and a rational attacker. We do 20 Monte Carlo trials, each with 15000 rounds. In Figure~\ref{fig:duality}, we observe that (i) the sensing team asymptotically approaches the game's equillbirum per \Cref{th:defender_regret}, and (ii) the duality gap, \ie the difference between the defenders' payoff and the attacker's payoff, decreases per \Cref{thm:epsilon_ne}. 

\paragraph{Impact of Network Optimization}
We demonstrate the effectiveness of network optimization per Algorithm~\ref{alg:neighbor} (\neighborsel).  We consider a $30\times 30$ environment. The sensors have FOV radius $r_i=8$, AOV $\theta_i=\pi/3$, communication range $c_i=16$, and heterogeneous bandwidth budgets $\alpha_i$ randomly chosen from $\{1, 2, 3\}$. We compare our framework with three benchmarks: \textit{(i) \actionsel + Nearest Neighbors:} Action selection follows Algorithm~\ref{alg:action}, while each sensor $i$ will always have the same closest peers as neighbors; \textit{(ii) \actionsel + Random Neighbors:} Action selection follows Algorithm~\ref{alg:action}, while each sensor $i$ will uniformly sample its neighbors from $\calM_i$ at each round; and \textit{(iii) \actionsel + All Possible Neighbors:} Action selection follows Algorithm~\ref{alg:action}, while each sensor $i$ has no bandwidth constraint and will always have all $\calM_i$ as neighbors (since \voc~is non-decreasing in $\calM_i$, this should achieve the best possible coverage). We conduct 20 Monte Carlo trials, each with 10000 rounds. From Figure~\ref{fig:coverage}, we observe: (i) our framework consistently outperforms the benchmarks with the same bandwidth budgets, and (ii) despite limited bandwidth, it is comparable to the best possible coverage with unlimited bandwidth.

\section{Conclusion} \label{sec:con}
In this paper, we proposed a distributed, game-theoretic framework for adversarial target coverage under limited communication. By modeling the interaction between the sensing team and the attacker as a zero-sum game with a combinatorial action space, we developed a decentralized algorithm that converges to an approximate NE while respecting communication constraints. Our approach enabled each agent to adaptively configure its communication neighborhood through an information-driven mechanism, allowing the system to scale efficiently. We established theoretical guarantees on convergence and quantified the near-optimality of the resulting strategies relative to the centralized game value. Through simulations, we showed that the proposed method converges to an approximate NE of the game and provides high coverage utility compared to other neighbor selection baselines.

\bibliographystyle{IEEEtran}
\bibliography{references}

\appendices
\section*{Appendix}
\subsection{Theoretical Proofs}
In the interest of space, we outline only the essential steps emphasizing the main technical ideas of the proofs.
\paragraph{Proof of \Cref{th:defender_regret}}
Consider a sequence of $\{b_t\}_{t\in[T]}$ that is selected by the attacker over $[T]$. 
Then, replacing $f$ with $f_t$ in~\cite[eq.~(14)]{xu2026self} (this preserves validity of the original argument), we have:
{%
\begin{align}\label{app-1}
    &\sum_{t=1}^T f_t(\calA_t) \geq (1-\curv_f)\, \sum_{t=1}^T f_t(\solopt)+ \kappa_f\, (1-\kappa_f)\, \rho(\kappa_{I}, \bar{\alpha}) \nonumber \\
    &\hspace{1.7cm}\times \sum_{i\in\calN} \sum_{t=1}^T \smi{a_{i,t}}{\calN_{i}^{\star}(\{a_{i,t}\}_{t\in [T]}; \alpha_i, \calM_i)} \nonumber\\
    &\hspace{2.1cm}- \Tilde{O}\left(|\calN|\sqrt{T\left(\bar{\alpha}^2\,|\bar{\calM}|\,+\,|\bar{\calV}|\right)}\right),
\end{align}}
\noindent where $\calN^*_i(\cdot)$ is the optimizer of \voc~\eqref{eq:SubMI}. Therefore,
{\allowdisplaybreaks%
\begin{align}\label{app-2}
    \calR^d_T &= \sum_{t=1}^T f_t(\solopt) - \sum_{t=1}^T f_t(\calA_t) \nonumber\\
    &\leq \curv_f\, \sum_{t=1}^T f_t(\solopt) - \kappa_f\, (1-\kappa_f)\, \rho(\kappa_{I}, \bar{\alpha}) \nonumber \\
    &\qquad\times \sum_{i\in\calN} \sum_{t=1}^T \smi{a_{i,t}}{\calN_{i}^{\star}(\{a_{i,t}\}_{t\in [T]}; \alpha_i, \calM_i)} \nonumber\\
    &\quad+ \Tilde{O}\left(|\calN|\sqrt{T\left(\bar{\alpha}^2\,|\bar{\calM}|\,+\,|\bar{\calV}|\right)}\right).
\end{align}}Since $0 \le f_t(\solopt) \le 1, \forall t$, \Cref{th:defender_regret} is proved. \qed

\paragraph{Proof of \Cref{lem:attacker_regret_bound}}
Since the attacker adopts \expthree, according to \cite[Theorem 3.1]{auer2002nonstochastic}, we have $\calR_T^a \leq \sqrt{2\,T\,|Y|\,\log{|Y|}}$, and thus \Cref{lem:attacker_regret_bound} holds. \qed

\paragraph{Proof of \Cref{thm:epsilon_ne}}
Consider the repeated-play dynamic induced by Algorithm~\ref{alg:main}, which generates a sequence of actions $\{(\mathcal{A}_t,b_t)\}_{t\in[T]}$. Since the payoff for a player is  linear in other player’s mixed strategy, the best fixed comparator in the regret definition is attained at an extreme point of the simplex, and thus can be taken to be a pure strategy. Theorem~\ref{th:defender_regret} and Lemma~\ref{lem:attacker_regret_bound} then bound the average regrets with respect to these best fixed comparators, as $\epsilon_1 \triangleq \mathcal{R}_T^d/T$ and $\epsilon_2 \triangleq \mathcal{R}_T^a/T$, respectively. Invoking \cite[Theorem 2]{farina2017regret} with $\epsilon_1$ and $\epsilon_2$, we get the result. \qed 

\subsection{The \actionsel and \neighborsel Algorithms}
\noindent Below, we outline Algorithms~\ref{alg:action} and \ref{alg:neighbor} in detail.
 \vspace{-3mm}
 \setlength{\textfloatsep}{3mm}
\begin{algorithm}[!h]
	\caption{\actionsel for Agent $i$ \cite{xu2026self}
	}
	\begin{algorithmic}[1]
		\REQUIRE Attacker's actions $(b_{1}...,b_{t-1})$, Number of time steps $T$, agent's action set $\calV_i$, objective function $f$
		\ENSURE \!Agent $i$'s distribution $p_{i,t}$ and action $a_{i, t}$.
		\medskip
            \STATE $\eta_i^a\gets\sqrt{2\log{|\calV_i|}\,/\,(|\calV_i|T)}$;
            \STATE $\hat{w}_{1}\gets\left[\hat{w}_{1,1}, \dots, \hat{w}_{|\calV_i|,1}\right]^\top$ with $\hat{w}_{a,1}=1, \forall a\in \calV_i$;
            \STATE $\distfsf_{i,1}\gets \hat{w}_1/\mathbf{1}^{\top} \hat{w}_1$
		
		\STATE \textbf{draw} action $a_{i,t}\in\calV_i$ \textbf{from} $\distfsf_{i,t}$;
        \textbf{receive} neighbors' actions $\{a_{j, t}\}_{j\in\calN_{i,t}}$;
		\STATE $r_{a_{i,t}, t}\gets f(\,a_{i,t}\,|\, \{a_{j, t}\}_{j\in\calN_{i,t}},b_t)$ and \\
  \textbf{normalize $r_{a_{i,t}, t}$ to} $[0,1]$;
            \STATE $\hat{r}_{a,\,t} \gets 1 - \frac{{\bf 1}(a_{i,t}\,=\,a)}{p_{a,t}}\left(1\,-\,r_{a_{i,t}, t}\right)$, $\forall a\in\calV_i$;
            \STATE $\hat{w}_{a,t+1}\gets \hat{w}_{a,t}\exp{(\eta_i^n \,\hat{r}_{a,t})}, \forall a\in\calV_i$;	
            \STATE \textbf{get} distribution $\distfsf_{i,t+1}\gets{\hat{w}_{t+1}}\,/\,{\mathbf{1}^{\top}\hat{w}_{t+1}}$;
	\end{algorithmic}\label{alg:action}
\end{algorithm}

 \vspace{-7mm}
 \setlength{\textfloatsep}{3mm}
\begin{algorithm}[!h]
	\caption{\neighborsel for Agent $i$ \cite{xu2026self}
	}
	\begin{algorithmic}[1]
		\REQUIRE Actions $a_{i,t}, (b_{1}...,b_{t-1})$, Number of time steps $T$, Agent $i$'s $\calM_i, \alpha_i$, and objective function $f$
		\ENSURE \!Agent $i$'s neighbors $\calN_{i,t}$ at each $t\in[T]$.
		\medskip
  		\STATE $\eta_i^n\gets\sqrt{2\log{|\calM_i|}\,/\,{(|\calM_i|T)}}$; 
		\STATE $z_{1}^{(k)}\gets\left[z_{1,\,1}^{(k)}, \dots, z_{\alpha_i,1}^{(k)}\right]^\top$; $z_{j,1}^{(k)}=1$, $\forall v\in \calM_i, \forall k\in[\alpha_i]$;  
            \FOR {$k = 1, \dots, \alpha_i$} 
            \STATE \textbf{get} distribution $\psi_{t}^{(k)}\gets z_t^{(k)}\,/\,{\|z_t^{(k)}\|_1}$; 
            \STATE \textbf{draw} agent $j_{t}^{(k)}\in\calM_i$ \textbf{from} $\psi_{t}^{(k)}$;
            \STATE \textbf{receive} action $a_{j_{t}^{(k)}, t}$ \textbf{from} $j_{t}^{(k)}$;
            \STATE $r_{j_{t}^{(k)}, t}\gets \smi{a_{i,t}}{\{a_{j_{t}^{(1)},t},\dots,a_{j_{t}^{(k)},t}\}} - $\\
            \hspace{1.3cm} $\smi{a_{i,t}}{\{a_{j_{t}^{(1)},t},\dots,a_{j_{t}^{(k-1)},t}\}}$ \\and
            \textbf{normalize $r_{j_{t}^{(k)}, t}$ to} $[0,1]$;
            \STATE $\hat{r}_{j,t}^{(k)} \gets 1 - \frac{{\bf 1}(j_{t}^{(k)}\,=\,j)}{q_{j,t}^{(k)}}\left(1\,-\,r_{j_{t}^{(k)},t}\right)$, $\forall j\in\calM_i$;
            \STATE $z_{j,t+1}^{(k)}\gets z_{j,\,t}^{(k)}\exp{(\eta_i^n \,\hat{r}_{j,t}^{(k)})}$, $\forall j\in\calM_i$;					\ENDFOR
                \STATE $\calN_{i,t}\gets\{j^{(k)}_{ t}\}_{k\in [\alpha_i]}$;
	\end{algorithmic}\label{alg:neighbor}
\end{algorithm}

\end{document}